%% file: susy01.proceedings.tex
\documentclass{ws-p8-50x6-00}

\begin{document}

\title{ Testing Theories of Fermion Masses }

\author{Stuart Raby}

\address{Department of Physics, The Ohio State University,
174 W. 18th Ave., Columbus, OH 43210\\ 
E-mail: raby@pacific.mps.ohio-state.edu}

\maketitle

\abstracts{The origin of quark and lepton masses is one of the 
outstanding problems of physics.  As the experimental data becomes 
more and more accurate, testing theories of fermion masses requires 
greater care.  In this talk we discuss a theoretical framework for 
testing those theories with a high energy desert.  
It is only with precision tests that we can hope to narrow the set 
of viable, beyond the standard model theories.}

\section{ Theoretical framework : top - down approach }
We consider theories in four dimensions with a high energy desert up to a physical threshold $M^*$ with $M_{pl} \ge M^* \ge M_G $, the GUT scale.  
Above $M^*$ we assume a fundamental theory  (such as string 
theory) from which the effective low energy theory is derived with
boundary conditions at $M^*$.  Below  $M^*$ we have two possibilities -
\begin{itemize}
\item $M^* \gg M_G$ and we have a SUSY GUT below.  In this case, the
relevant boundary conditions for low energy phenomenology are defined at 
$M_G$.
\item  Or we have the minimal SUSY spectrum below $M^*$ with relevant boundary
conditions there.
\end{itemize}
The latter possibility is the default when talking of string unification.
I focus on the former possibility in the rest of this talk.

\subsection{SUSY GUT with boundary conditions at $M_G$}
In this case we have three types of boundary conditions specifying the parameters for  gauge couplings, Yukawa couplings and soft SUSY breaking
at $M_G$.
\begin{itemize}
\item {\bf Gauge couplings} \hspace{.4in}   $\alpha_i = \alpha_G  \;\;{\rm for} \; i = 1, \cdots, 3$.

 This is a tree level relation in a GUT. A good fit to the low energy data already requires one loop threshold corrections; hence these corrections, predominantly from the GUT breaking and Higgs sectors of the theory,
are observable and thus constrain GUTs.

\item {\bf Yukawa couplings}   
  
Schematically we assume
\begin{eqnarray}
Y =&  \left(\begin{array}{ccc}  0 & \epsilon^\prime  & 0 \\
                           \epsilon^\prime  &  \epsilon  &   \epsilon \\
                      0  &  \epsilon  & 1 \end{array} \right)  &   \nonumber   \end{eqnarray}
where $\epsilon,\;  \epsilon^\prime/\epsilon$ are small parameters.
One generally (with or without GUTs)
assumes some family symmetry; whether it is a permutation, U(1), or non-abelian
SU(2), SU(3), or discrete non-abelian symmetry D(3).
Family symmetries are needed for several reasons:
\begin{itemize}
 \item they reduce the number of fundamental parameters;

\item they can explain the hierarchy of fermion masses;

\item they can suppress flavor violation, such as $\mu \rightarrow e \gamma$;

\item they are necessary to suppress dimension 5 baryon number violating operators responsible for nucleon decay, i.e. they explain the small Yukawa-like
couplings appearing in these dimension five operators. Or they can be used
(in theories without GUTs) to forbid, with U(1) or baryon parity, dimension
5 baryon number violating operators all together;   

\item and they can eliminate any other undesirable effects of Planck scale physics.
\end{itemize}

In general, there are three Yukawa matrices given by

\begin{eqnarray}
Y_k =&  \left(\begin{array}{ccc}  0 & D_k \; \epsilon'  & 0 \\
   \hat D_k \; \epsilon'  &  C_k \; \epsilon  &  B_k \; \epsilon \\
   0  & \hat B_k \; \epsilon  & A_k \end{array} \right) &   \nonumber   \end{eqnarray}
with ($k =  u, \; d, \; e $).  The parameters $A, \; B,\; \hat B,\; C,\;
D, \; \hat D$ are arbitrary, order one complex coefficients; fit
to the data.

\item {\bf SUSY breaking}

For example, in the CMSSM we have $ m_0, \;\; M_{1/2}, \;\; A_0,  \;\; \tan\beta, \;\; sign(\mu)$.   In other SUSY breaking
scenaria (such as, GMSB, AMSB, $\tilde g$MSB, ... ) we have a different set 
of SUSY breaking parameters.  The main point is that simple mechanisms for
SUSY breaking have a small number of arbitrary parameters; hence they
can be predictive.
\end{itemize}

\section{ Precision Electroweak Data : including fermion masses
and mixing angles }
The precision electroweak data for testing theories beyond the standard model
includes, in addition to the electroweak data used for testing the standard model, fermion masses and mixing angles.   
\begin{itemize}
\item {SM (CMSSM) Precision Data}

$$\begin{array}{ll} M_Z =  91.188 \;\; (0.091) &  M_W =  80.419 \;\; (0.080) \\
 G_{\mu}\cdot 10^5 =  1.1664  \;\; (0.0012) & \alpha_{EM}^{-1} = 137.04  \;\; (0.14)\\ 
\alpha_s(M_Z) =  0.1181  \;\; (0.002) & \rho_{new}\cdot 10^3 =   -0.20  \;\; (1.1) \\
B(b \rightarrow s \gamma) \cdot 10^{4} =  2.96  \;\; (0.35) &
\end{array}$$

\item{BSM Precision Data}

$$\begin{array}{ll} M_t =  174.3  \;\; (5.1)  & m_b(m_b) =  4.20 \;\; (0.20) \\  
 M_b - M_c  =   3.4  \;\; (0.2) &  m_c(m_c) = 1.25 \;\; (0.10) \\
 m_s(2 {\rm GeV}) = 110  \;\; (35) & m_d/m_s  =  0.050  \;\; (0.010) \\ 
Q =   22.7  \;\; (0.8) &  \\  
M_{\tau} =  1.777  \;\; (0.0018) &  M_{\mu} =   0.10566  \;\; (0.00011) \\   
M_e \cdot 10^3  =  0.5110  \;\; (0.00051) &  \\
|V_{us}|  = 0.2196  \;\; (0.0023) &  |V_{cb}| =   0.0402  \;\; (0.0019) \\     
 |V_{ub}/V_{cb}| =  0.090 \;\; (0.025)  & \\
\epsilon_K  \cdot 10^3  =   2.28  \;\; (0.013) & \hat B_K = 0.85 \;\; (0.15)   
\end{array}$$   
\end{itemize}

I have included only a subset of the SM precision data.  This is the subset
we have used in our analysis~\cite{bcrw}.  Instead of the complete set of
standard model data we have added the thirteen observables
associated with charged fermion masses and mixing angles.
With regards to the BSM precision data, $M (m)$ refer to pole ($\overline{MS}$ running) masses.  $M_b - M_c$ has smaller theoretical uncertainties than
$m_c(m_c)$. This is because one can use HQEFT to relate the difference to $M_B - M_D$ with renormalon effects cancelling.  Hence we have used the former rather than the latter.     $Q$ is the Kaplan- Manohar- Leutwyler ellipse parameter defined by $$ \frac{m_u^2}{m_d^2} + \frac{1}{Q^2} \frac{m_s^2}{m_d^2} = 1 .$$  Note, this is the best known relation among
the three light quark masses.   The experimental results $X_k^{exp}$ ($\sigma_k$)
are given above.   These are for the most part taken from the PDG 2000. 
We have however used only the lower half of the PDG range for $m_s$;
as suggested by recent lattice results.  Finally, in those
cases where the experimental uncertainty is less than 0.1\% we assume a numerical error of 0.1\% for $\sigma_k$.   We then define a $\chi^2$ function
given by  $$\chi^2 =  \sum_{k=1}^{N} (X_k^{exp} - X_k^{th})^2/\sigma_k^2 .$$

\subsection{ Global $\chi^2$ fits }
We fit the data by minimizing the $\chi^2$ function using the program
Minuit.  For the most part,
the parameters we vary are defined at the GUT scale, except for $\mu$
and $\tan\beta$ defined at $M_Z$.   We choose boundary conditions at $M_G$
given by

$\bullet$  $\alpha_1 = \alpha_2 \equiv \alpha_G$, $\alpha_3 = \alpha_G \; (1 + 
\epsilon_3)$  (where $\epsilon_3$, parametrizing the one loop threshold
corrections to gauge coupling unification, is a free parameter);

$\bullet$  a model dependent parametrization of Yukawa couplings; and

$\bullet$  soft SUSY breaking parameters which we take to be  
universal squark and slepton masses -  $m_{16}$, universal gaugino masses - $M_{1/2}$, universal cubic scalar coupling - $A_0$, non-universal  $H_u,  \;\; H_d$ masses, $\mu$ and $\tan\beta(M_Z)$.

We use two loop renormalization group running from  $M_G$ to $M_Z$
within the MSSM.  We include one loop threshold corrections to fermion masses
and electroweak observables at $M_Z$.   Below $M_Z$ we use 3 loop QCD + 1 loop QED running.    Electroweak symmetry breaking is obtained self-consistently
using the one loop improved Higgs potential, including $m_t^4$ and
$m_b^4$ corrections with an effective 2 Higgs doublet model defined below $M_{stop}$~\cite{cqw}.   Finally the $\chi^2$ function includes (for three
families) 20 low energy observables.   

Our analysis is in the same spirit as the global fits of the standard model (or
MSSM) to precision electroweak data~\cite{deboer}, although we use only a subset of the electroweak data. 
In the rest of this talk I focus on our recent analysis~\cite{bdr},
including only the third generation quark and lepton masses.

\section{ SO(10) Yukawa unification for third generation and threshold corrections at $M_Z$ }

In the simplest version of $SO_{10}$ the Yukawa couplings for the third generation are all equal at $M_G$, i.e. $\lambda_b = \lambda_t = 
\lambda_{\tau} = \lambda_{\nu_\tau} = \lambda$.
The low energy values of the top, bottom and $\tau$ couplings depend
solely on the value of $M_G$, $\alpha_G$ and $\lambda$.  The first
two are already fixed by gauge coupling unification, hence we need only
one low energy observable to fix $\lambda$.   We use the ratio
$m_b/m_{\tau}$.   We then use  $m_\tau = \lambda_\tau \frac{v}{\sqrt{2}} cos\beta$ to fix the value of  $\beta $; $v$ is fixed by electroweak data.   Finally we obtain one prediction for
$ {m_t = \lambda_t \frac{v}{\sqrt{2}} sin\beta \sim 170 \pm 20 \; {\rm GeV}}$
~\cite{so10yukunif}.  

\subsection{Weak scale threshold corrections -  large $tan\beta$ }
This beautiful result is however spoiled by potentially large weak scale threshold corrections proportional to $\tan\beta$~\cite{threshcor}.
By far the largest corrections come from SUSY mass insertions given by
$\delta m_b \sim  \delta m_b^{\tilde g}  +  \delta m_b^{\tilde \chi} $
where
$\delta m_b^{\tilde g}  \propto \frac{\mu m_{\tilde g} tan\beta}{m_{\tilde b}^2}$
and
$\delta m_b^{\tilde \chi}  \propto \frac{\mu A_t tan\beta}{m_{\tilde t}^2} $.
These corrections can be as large as $\sim  50$ \%; however good fits require $\delta m_b < -2$\%.    

Note that when $\delta m_b^{\tilde g} > 0$ ($\mu > 0$, {\it in our convention}), then $\delta m_b^{\tilde \chi} < 0$. 
Recent data favors $\mu > 0$.  For example, the process 
$b \rightarrow s \gamma$ is obtained
by similar graphs with a photon insertion.  
In this case the chargino term typically dominates and 
has opposite sign to the SM and charged Higgs contributions for $\mu > 0$, thus reducing the branching ratio.  This is a good thing, since the standard model
plus charged Higgs contribution is typically larger than the experimental
value.  Thus {\bf $\mu < 0$ is problematic.}  Also the sign of the SUSY contribution to the anomalous magnetic moment of the muon is correlated with the sign of $\mu$.  Recent BNL results favor $\mu > 0$~\cite{bnl}.  

Pierce et al.~\cite{pierceetal} and Baer et al.~\cite{baeretal} find Yukawa unification only when
$\delta m_b^{\tilde g}  < 0$ and hence poor fits for $b \rightarrow s \gamma$. 
On the other hand we find good fits for $\delta m_b^{\tilde g} > 0$~\cite{bcrw}.
What is going on?

In the past, we have neglected the small ``wavefunction renormalisation" corrections to $\delta m_b$.  These corrections, $\delta m_b^{log}$, are not proportional to $\tan\beta$ nor to $A_t$ or $m_{\tilde g}$.  They are only logarithmically dependent on SUSY particle masses.  They are however positive and about 6\%.  They thus have to be cancelled in order
to find good fits to the bottom mass.  Note, the corrections to the top and
$\tau$ mass are small.   In the present analysis~\cite{bdr} we include all one loop threshold corrections~\cite{pierceetal} to quark, charged lepton masses 
and electroweak observables. 

{\em We find Yukawa unification possible only in a narrow
region of soft SUSY breaking parameter space.}
 
In this region of parameter space with large $\tan\beta$ we need
large $m_{16}, \; A_0$ and
smaller $\mu, \; M_{1/2}$.  A large negative value of $A_0$ serves two
purposes:
\begin{itemize}
\item  it enhances the chargino contribution and
\item  it induces large splitting between $\tilde t_{1} - \tilde t_2$.
\end{itemize}  Both effects go towards enhancing 
$\delta m_b^{\tilde \chi}$ compared to $\delta m_b^{\tilde g}  + 
\delta m_b^{log}$.  In addition large values of $m_{16}$ enable the largest
possible splitting of $\tilde t_{1,\; 2}$ while maintaining a positive
stop mass squared.  

This range of parameters is however consistent
with electroweak symmetry breaking only if the $H_u, \; H_d$ masses
are split.   We have considered two possible ways of splitting the
Higgs masses we refer to as 
$D_X$ term splitting  and  universal scalar masses.

{\bf $D_X$ splitting}

In this case we have
$ m_{(H_u, H_d)}^2 =  m_{10}^2 \mp 2 \; D_X$,
$ m_{(Q,\; \bar u,\; \bar e)}^2 =  m_{16}^2 + D_X$ and 
$ m_{(\bar d,\; L)}^2 =  m_{16}^2 - 3 D_X$.
Unfortunately, since $D_X > 0$ we find
$ m_{\tilde b}^2  <<  m_{\tilde t}^2$  which is {\em bad}.

{\bf Universal scalar masses}

In the other case we have 
$ m_{(H_u, H_d)}^2 =  m_{10}^2 \mp 2 \; \Delta m_H^2$, and 
$ m_{(Q,\; \bar u,\; \bar e)}^2 = m_{(\bar d,\; L)}^2 =  m_{16}^2$.  
We now find $ m_{\tilde t}^2  <<  m_{\tilde b}^2$ which
is just what is needed, i.e.  {\em good}.  

In figure 1 we give contours of constant $\chi^2$ as a function
of $\mu, \; M_{1/2}$ with $m_{16} = 1500$ and $2000$ GeV.
All other parameters have been varied in order to minimize $\chi^2$.
In figure 2 we show $m_b(m_b)$ and $\delta m_b$ contours as a function of
$\mu, \; M_{1/2}$ with $m_{16} = 2000$ GeV.   
By comparing the two figures, it is clear that $\chi^2$ is smallest
when $\delta m_b < -2$\%.  The best fits are obtained with $A_0 \sim - 1.9 \; m_{16}$ and $m_{10} \sim 1.4 \; m_{16}$.

\begin{figure}[t]
\begin{center}

\input{chi2.pstex_t}

\caption{$\chi^2$ contours for $m_{16} = 1500$ GeV (Left) and $m_{16} = 2000$
 GeV (Right).  The shaded region is excluded by the chargino mass limit
$m_{\chi^+} > 103$ GeV.}
\label{figure:chi2}
\end{center}
\end{figure}
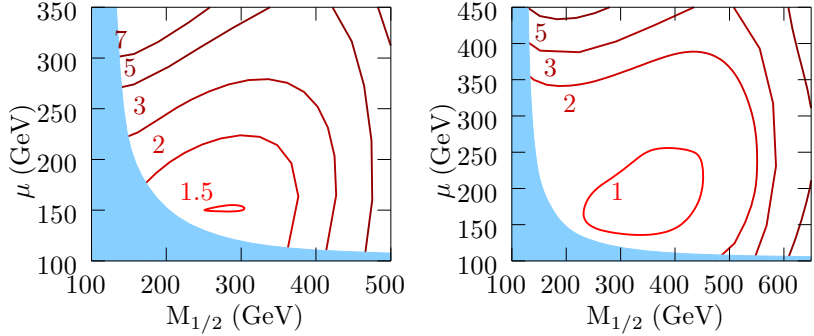

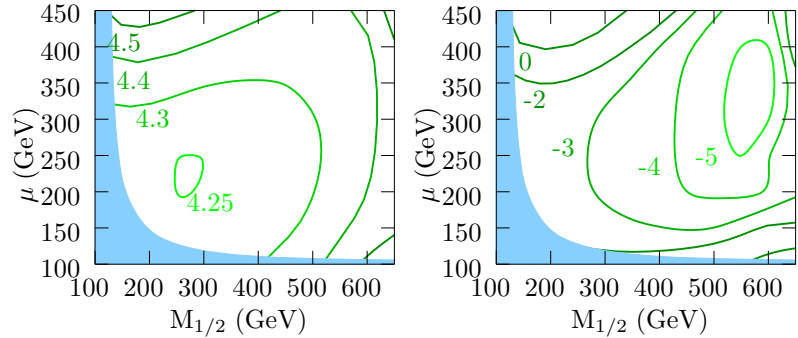
\begin{figure}[t]
\begin{center}

\input{m_b_and_corr.pstex_t}

\caption{Contours of constant $m_b(m_b)$[GeV] (Left) and $\delta m_b$ in \% (Right)
for $m_{16} = 2000$ GeV.}
\label{figure:m_b_and_corr}
\end{center}
\end{figure}

In a more complete theory including all three generations, we would be
able to calculate the anomalous magnetic moment of the muon $a_\mu^{NEW}$
and the proton lifetime.  Here we show results obtained in a particular
3 family model (see Blazek, S.R. and Tobe~\cite{bcrw}).  
We find $a_\mu^{NEW}$ is less than the recent BNL measurement
$a_\mu^{NEW} = 43 \pm 16 \times 10^{-10}$.~\cite{bnl}  In fact $a_\mu^{NEW}$ decreases with increasing $m_{16}$.  With a minimum value of $m_{16} \sim 1200$ GeV, necessary for reasonable fits, we find $a_\mu^{NEW} \leq 16 \times 10^{-10}$.

The proton lifetime, on the other hand, increases as $m_{16}$ increases.
It can be consistent with the most recent Super K preliminary results~\cite{superk}, but
only if we allow for the three quark matrix element $\beta$
between a proton state
and the vacuum to be of order 0.5 $\times \beta_{lattice}$~\cite{lattice}.

\section{ Conclusion }
In conclusion, theories beyond the standard model are strongly constrained
by precision electroweak data, which now includes fermion masses and mixing angles.
As an example, we have discussed SO(10) Yukawa unification.  We find
that good fits constrain the soft SUSY breaking parameters to a narrow region
given by $m_{16} \geq 1200$ GeV,  $A_0 \sim - 1.9 \; m_{16}$, and $m_{10} \sim
1.4 \; m_{16}$.   This then implies significant predictions for Higgs and SUSY particle masses.   See the talk of R. Derm\' \i \v sek in these proceedings
for further details.

\begin{center}
\epsfig{file=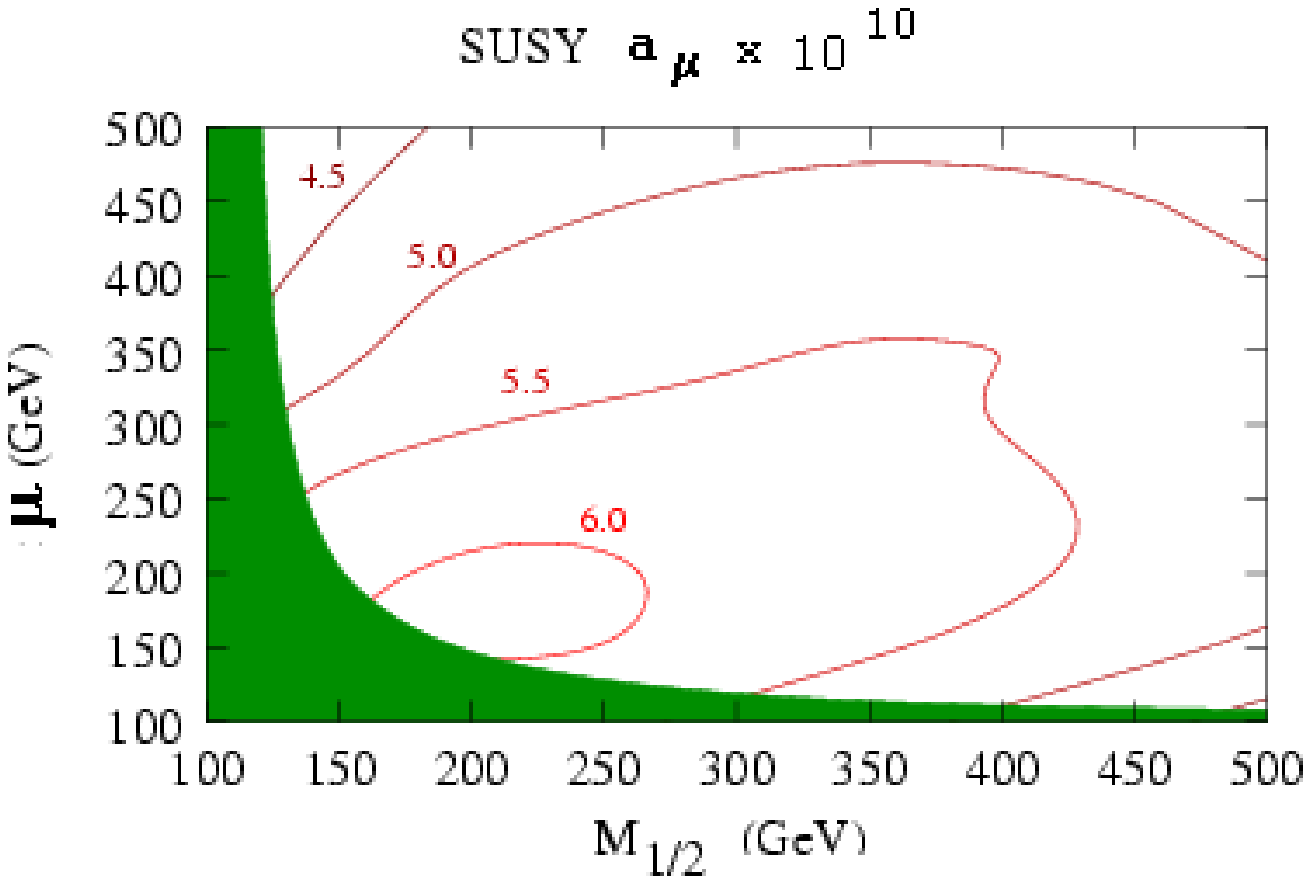,angle=0,width=3.7in}
\end{center}

\begin{center}
\epsfig{file=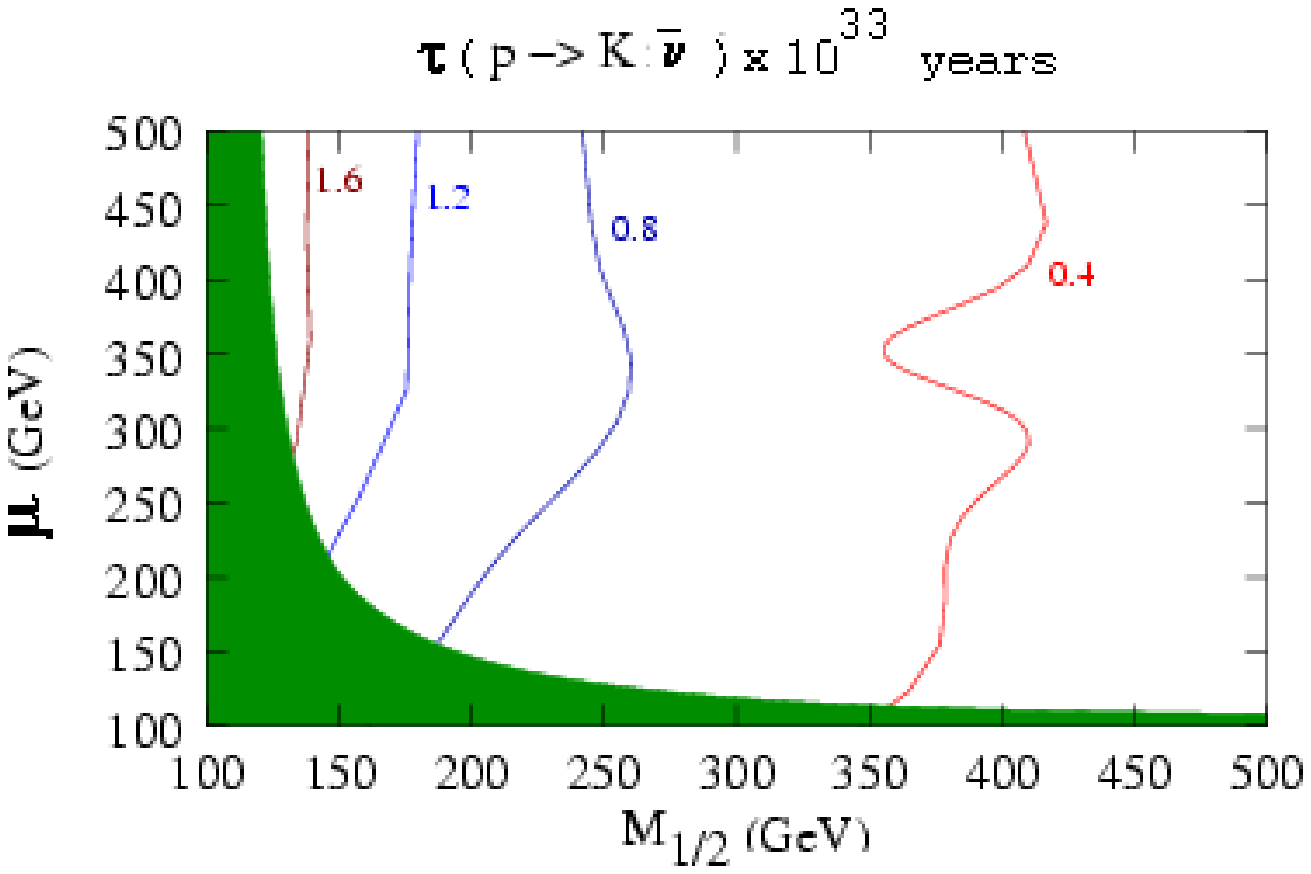,angle=0,width=3.7in}
\end{center}

\noindent
{\small Figure 3:  $a_\mu^{NEW}$ and the proton lifetime plotted as a function
of $\mu, \; M_{1/2}$ for $m_{16} = 2000$ GeV.}

\end{document}

%% file: chi2.pstex_t
\begin{picture}(0,0)%
\includegraphics{chi2.pstex}%
\end{picture}%
\setlength{\unitlength}{3158sp}%
\begingroup\makeatletter\ifx\SetFigFont\undefined%
\gdef\SetFigFont#1#2#3#4#5{%
  \reset@font\fontsize{#1}{#2pt}%
  \fontfamily{#3}\fontseries{#4}\fontshape{#5}%
  \selectfont}%
\fi\endgroup%
\begin{picture}(6342,2524)(556,-3511)
\put(1651,-1861){\makebox(0,0)[b]{\smash{\SetFigFont{10}{12.0}{\familydefault}{\mddefault}{\updefault}\special{ps: gsave .69 0 0 setrgbcolor}3\special{ps: grestore}
}}}
\put(1576,-1561){\makebox(0,0)[b]{\smash{\SetFigFont{10}{12.0}{\familydefault}{\mddefault}{\updefault}\special{ps: gsave .56 0 0 setrgbcolor}5\special{ps: grestore}
}}}
\put(2101,-2536){\makebox(0,0)[b]{\smash{\SetFigFont{10}{12.0}{\familydefault}{\mddefault}{\updefault}\special{ps: gsave 1 0 0 setrgbcolor}1.5\special{ps: grestore}
}}}
\put(1801,-2161){\makebox(0,0)[b]{\smash{\SetFigFont{10}{12.0}{\familydefault}{\mddefault}{\updefault}\special{ps: gsave .82 0 0 setrgbcolor}2\special{ps: grestore}
}}}
\put(1501,-1336){\makebox(0,0)[b]{\smash{\SetFigFont{10}{12.0}{\familydefault}{\mddefault}{\updefault}\special{ps: gsave .56 0 0 setrgbcolor}7\special{ps: grestore}
}}}
\put(4876,-1561){\makebox(0,0)[b]{\smash{\SetFigFont{10}{12.0}{\familydefault}{\mddefault}{\updefault}\special{ps: gsave .69 0 0 setrgbcolor}3\special{ps: grestore}
}}}
\put(5026,-1841){\makebox(0,0)[b]{\smash{\SetFigFont{10}{12.0}{\familydefault}{\mddefault}{\updefault}\special{ps: gsave .82 0 0 setrgbcolor}2\special{ps: grestore}
}}}
\put(4801,-1286){\makebox(0,0)[b]{\smash{\SetFigFont{10}{12.0}{\familydefault}{\mddefault}{\updefault}\special{ps: gsave .56 0 0 setrgbcolor}5\special{ps: grestore}
}}}
\put(5401,-2536){\makebox(0,0)[b]{\smash{\SetFigFont{10}{12.0}{\familydefault}{\mddefault}{\updefault}\special{ps: gsave 1 0 0 setrgbcolor}1\special{ps: grestore}
}}}
\put(1813,-3261){\makebox(0,0)[b]{\smash{\SetFigFont{10}{12.0}{\familydefault}{\mddefault}{\updefault}\special{ps: gsave 0 0 0 setrgbcolor}200\special{ps: grestore}
}}}
\put(2401,-3261){\makebox(0,0)[b]{\smash{\SetFigFont{10}{12.0}{\familydefault}{\mddefault}{\updefault}\special{ps: gsave 0 0 0 setrgbcolor}300\special{ps: grestore}
}}}
\put(2988,-3261){\makebox(0,0)[b]{\smash{\SetFigFont{10}{12.0}{\familydefault}{\mddefault}{\updefault}\special{ps: gsave 0 0 0 setrgbcolor}400\special{ps: grestore}
}}}
\put(3576,-3261){\makebox(0,0)[b]{\smash{\SetFigFont{10}{12.0}{\familydefault}{\mddefault}{\updefault}\special{ps: gsave 0 0 0 setrgbcolor}500\special{ps: grestore}
}}}
\put(1001,-2716){\makebox(0,0)[b]{\smash{\SetFigFont{10}{12.0}{\familydefault}{\mddefault}{\updefault}\special{ps: gsave 0 0 0 setrgbcolor}150\special{ps: grestore}
}}}
\put(1001,-2318){\makebox(0,0)[b]{\smash{\SetFigFont{10}{12.0}{\familydefault}{\mddefault}{\updefault}\special{ps: gsave 0 0 0 setrgbcolor}200\special{ps: grestore}
}}}
\put(1001,-1919){\makebox(0,0)[b]{\smash{\SetFigFont{10}{12.0}{\familydefault}{\mddefault}{\updefault}\special{ps: gsave 0 0 0 setrgbcolor}250\special{ps: grestore}
}}}
\put(1001,-1521){\makebox(0,0)[b]{\smash{\SetFigFont{10}{12.0}{\familydefault}{\mddefault}{\updefault}\special{ps: gsave 0 0 0 setrgbcolor}300\special{ps: grestore}
}}}
\put(1001,-1122){\makebox(0,0)[b]{\smash{\SetFigFont{10}{12.0}{\familydefault}{\mddefault}{\updefault}\special{ps: gsave 0 0 0 setrgbcolor}350\special{ps: grestore}
}}}
\put(1226,-3261){\makebox(0,0)[b]{\smash{\SetFigFont{10}{12.0}{\familydefault}{\mddefault}{\updefault}\special{ps: gsave 0 0 0 setrgbcolor}100\special{ps: grestore}
}}}
\put(1001,-3115){\makebox(0,0)[b]{\smash{\SetFigFont{10}{12.0}{\familydefault}{\mddefault}{\updefault}\special{ps: gsave 0 0 0 setrgbcolor}100\special{ps: grestore}
}}}
\put(2401,-3511){\makebox(0,0)[b]{\smash{\SetFigFont{10}{12.0}{\familydefault}{\mddefault}{\updefault}\special{ps: gsave 0 0 0 setrgbcolor}$\rm M_{1/2} \; ( GeV )$\special{ps: grestore}
}}}
\put(751,-2118){\rotatebox{90.0}{\makebox(0,0)[b]{\smash{\SetFigFont{10}{12.0}{\familydefault}{\mddefault}{\updefault}\special{ps: gsave 0 0 0 setrgbcolor}$\mu \rm \; ( GeV )$\special{ps: grestore}
}}}}
\put(4953,-3261){\makebox(0,0)[b]{\smash{\SetFigFont{10}{12.0}{\familydefault}{\mddefault}{\updefault}\special{ps: gsave 0 0 0 setrgbcolor}200\special{ps: grestore}
}}}
\put(5380,-3261){\makebox(0,0)[b]{\smash{\SetFigFont{10}{12.0}{\familydefault}{\mddefault}{\updefault}\special{ps: gsave 0 0 0 setrgbcolor}300\special{ps: grestore}
}}}
\put(5807,-3261){\makebox(0,0)[b]{\smash{\SetFigFont{10}{12.0}{\familydefault}{\mddefault}{\updefault}\special{ps: gsave 0 0 0 setrgbcolor}400\special{ps: grestore}
}}}
\put(6235,-3261){\makebox(0,0)[b]{\smash{\SetFigFont{10}{12.0}{\familydefault}{\mddefault}{\updefault}\special{ps: gsave 0 0 0 setrgbcolor}500\special{ps: grestore}
}}}
\put(6662,-3261){\makebox(0,0)[b]{\smash{\SetFigFont{10}{12.0}{\familydefault}{\mddefault}{\updefault}\special{ps: gsave 0 0 0 setrgbcolor}600\special{ps: grestore}
}}}
\put(5701,-3511){\makebox(0,0)[b]{\smash{\SetFigFont{10}{12.0}{\familydefault}{\mddefault}{\updefault}\special{ps: gsave 0 0 0 setrgbcolor}$\rm M_{1/2} \; ( GeV )$\special{ps: grestore}
}}}
\put(4301,-2830){\makebox(0,0)[b]{\smash{\SetFigFont{10}{12.0}{\familydefault}{\mddefault}{\updefault}\special{ps: gsave 0 0 0 setrgbcolor}150\special{ps: grestore}
}}}
\put(4301,-2546){\makebox(0,0)[b]{\smash{\SetFigFont{10}{12.0}{\familydefault}{\mddefault}{\updefault}\special{ps: gsave 0 0 0 setrgbcolor}200\special{ps: grestore}
}}}
\put(4301,-2261){\makebox(0,0)[b]{\smash{\SetFigFont{10}{12.0}{\familydefault}{\mddefault}{\updefault}\special{ps: gsave 0 0 0 setrgbcolor}250\special{ps: grestore}
}}}
\put(4301,-1976){\makebox(0,0)[b]{\smash{\SetFigFont{10}{12.0}{\familydefault}{\mddefault}{\updefault}\special{ps: gsave 0 0 0 setrgbcolor}300\special{ps: grestore}
}}}
\put(4301,-1691){\makebox(0,0)[b]{\smash{\SetFigFont{10}{12.0}{\familydefault}{\mddefault}{\updefault}\special{ps: gsave 0 0 0 setrgbcolor}350\special{ps: grestore}
}}}
\put(4301,-1407){\makebox(0,0)[b]{\smash{\SetFigFont{10}{12.0}{\familydefault}{\mddefault}{\updefault}\special{ps: gsave 0 0 0 setrgbcolor}400\special{ps: grestore}
}}}
\put(4301,-1122){\makebox(0,0)[b]{\smash{\SetFigFont{10}{12.0}{\familydefault}{\mddefault}{\updefault}\special{ps: gsave 0 0 0 setrgbcolor}450\special{ps: grestore}
}}}
\put(4301,-3115){\makebox(0,0)[b]{\smash{\SetFigFont{10}{12.0}{\familydefault}{\mddefault}{\updefault}\special{ps: gsave 0 0 0 setrgbcolor}100\special{ps: grestore}
}}}
\put(4051,-2118){\rotatebox{90.0}{\makebox(0,0)[b]{\smash{\SetFigFont{10}{12.0}{\familydefault}{\mddefault}{\updefault}\special{ps: gsave 0 0 0 setrgbcolor}$\mu \rm \; ( GeV )$\special{ps: grestore}
}}}}
\put(4526,-3261){\makebox(0,0)[b]{\smash{\SetFigFont{10}{12.0}{\familydefault}{\mddefault}{\updefault}\special{ps: gsave 0 0 0 setrgbcolor}100\special{ps: grestore}
}}}
\end{picture}

%% file: m_b_and_corr.pstex_t
\begin{picture}(0,0)%
\includegraphics{m_b_and_corr.pstex}%
\end{picture}%
\setlength{\unitlength}{3158sp}%
\begingroup\makeatletter\ifx\SetFigFont\undefined%
\gdef\SetFigFont#1#2#3#4#5{%
  \reset@font\fontsize{#1}{#2pt}%
  \fontfamily{#3}\fontseries{#4}\fontshape{#5}%
  \selectfont}%
\fi\endgroup%
\begin{picture}(6192,2535)(556,-3511)
\put(1501,-1336){\makebox(0,0)[b]{\smash{\SetFigFont{10}{12.0}{\familydefault}{\mddefault}{\updefault}\special{ps: gsave 0 .56 0 setrgbcolor}4.5\special{ps: grestore}
}}}
\put(1726,-1936){\makebox(0,0)[b]{\smash{\SetFigFont{10}{12.0}{\familydefault}{\mddefault}{\updefault}\special{ps: gsave 0 .82 0 setrgbcolor}4.3\special{ps: grestore}
}}}
\put(1576,-1636){\makebox(0,0)[b]{\smash{\SetFigFont{10}{12.0}{\familydefault}{\mddefault}{\updefault}\special{ps: gsave 0 .69 0 setrgbcolor}4.4\special{ps: grestore}
}}}
\put(2176,-2591){\makebox(0,0)[b]{\smash{\SetFigFont{10}{12.0}{\familydefault}{\mddefault}{\updefault}\special{ps: gsave 0 1 0 setrgbcolor}4.25\special{ps: grestore}
}}}
\put(4651,-1486){\makebox(0,0)[b]{\smash{\SetFigFont{10}{12.0}{\familydefault}{\mddefault}{\updefault}\special{ps: gsave 0 .56 0 setrgbcolor}0\special{ps: grestore}
}}}
\put(4726,-1786){\makebox(0,0)[b]{\smash{\SetFigFont{10}{12.0}{\familydefault}{\mddefault}{\updefault}\special{ps: gsave 0 .56 0 setrgbcolor}-2\special{ps: grestore}
}}}
\put(6076,-2236){\makebox(0,0)[b]{\smash{\SetFigFont{10}{12.0}{\familydefault}{\mddefault}{\updefault}\special{ps: gsave 0 1 0 setrgbcolor}-5\special{ps: grestore}
}}}
\put(5626,-2311){\makebox(0,0)[b]{\smash{\SetFigFont{10}{12.0}{\familydefault}{\mddefault}{\updefault}\special{ps: gsave 0 .82 0 setrgbcolor}-4\special{ps: grestore}
}}}
\put(4951,-2161){\makebox(0,0)[b]{\smash{\SetFigFont{10}{12.0}{\familydefault}{\mddefault}{\updefault}\special{ps: gsave 0 .69 0 setrgbcolor}-3\special{ps: grestore}
}}}
\put(1226,-3261){\makebox(0,0)[b]{\smash{\SetFigFont{10}{12.0}{\familydefault}{\mddefault}{\updefault}\special{ps: gsave 0 0 0 setrgbcolor}100\special{ps: grestore}
}}}
\put(1653,-3261){\makebox(0,0)[b]{\smash{\SetFigFont{10}{12.0}{\familydefault}{\mddefault}{\updefault}\special{ps: gsave 0 0 0 setrgbcolor}200\special{ps: grestore}
}}}
\put(2080,-3261){\makebox(0,0)[b]{\smash{\SetFigFont{10}{12.0}{\familydefault}{\mddefault}{\updefault}\special{ps: gsave 0 0 0 setrgbcolor}300\special{ps: grestore}
}}}
\put(2507,-3261){\makebox(0,0)[b]{\smash{\SetFigFont{10}{12.0}{\familydefault}{\mddefault}{\updefault}\special{ps: gsave 0 0 0 setrgbcolor}400\special{ps: grestore}
}}}
\put(2935,-3261){\makebox(0,0)[b]{\smash{\SetFigFont{10}{12.0}{\familydefault}{\mddefault}{\updefault}\special{ps: gsave 0 0 0 setrgbcolor}500\special{ps: grestore}
}}}
\put(3362,-3261){\makebox(0,0)[b]{\smash{\SetFigFont{10}{12.0}{\familydefault}{\mddefault}{\updefault}\special{ps: gsave 0 0 0 setrgbcolor}600\special{ps: grestore}
}}}
\put(2401,-3511){\makebox(0,0)[b]{\smash{\SetFigFont{10}{12.0}{\familydefault}{\mddefault}{\updefault}\special{ps: gsave 0 0 0 setrgbcolor}$\rm M_{1/2} \; ( GeV )$\special{ps: grestore}
}}}
\put(1001,-3115){\makebox(0,0)[b]{\smash{\SetFigFont{10}{12.0}{\familydefault}{\mddefault}{\updefault}\special{ps: gsave 0 0 0 setrgbcolor}100\special{ps: grestore}
}}}
\put(1001,-2830){\makebox(0,0)[b]{\smash{\SetFigFont{10}{12.0}{\familydefault}{\mddefault}{\updefault}\special{ps: gsave 0 0 0 setrgbcolor}150\special{ps: grestore}
}}}
\put(1001,-2546){\makebox(0,0)[b]{\smash{\SetFigFont{10}{12.0}{\familydefault}{\mddefault}{\updefault}\special{ps: gsave 0 0 0 setrgbcolor}200\special{ps: grestore}
}}}
\put(1001,-2261){\makebox(0,0)[b]{\smash{\SetFigFont{10}{12.0}{\familydefault}{\mddefault}{\updefault}\special{ps: gsave 0 0 0 setrgbcolor}250\special{ps: grestore}
}}}
\put(1001,-1976){\makebox(0,0)[b]{\smash{\SetFigFont{10}{12.0}{\familydefault}{\mddefault}{\updefault}\special{ps: gsave 0 0 0 setrgbcolor}300\special{ps: grestore}
}}}
\put(1001,-1691){\makebox(0,0)[b]{\smash{\SetFigFont{10}{12.0}{\familydefault}{\mddefault}{\updefault}\special{ps: gsave 0 0 0 setrgbcolor}350\special{ps: grestore}
}}}
\put(1001,-1407){\makebox(0,0)[b]{\smash{\SetFigFont{10}{12.0}{\familydefault}{\mddefault}{\updefault}\special{ps: gsave 0 0 0 setrgbcolor}400\special{ps: grestore}
}}}
\put(751,-2118){\rotatebox{90.0}{\makebox(0,0)[b]{\smash{\SetFigFont{10}{12.0}{\familydefault}{\mddefault}{\updefault}\special{ps: gsave 0 0 0 setrgbcolor}$\mu \rm \; ( GeV )$\special{ps: grestore}
}}}}
\put(1001,-1111){\makebox(0,0)[b]{\smash{\SetFigFont{10}{12.0}{\familydefault}{\mddefault}{\updefault}\special{ps: gsave 0 0 0 setrgbcolor}450\special{ps: grestore}
}}}
\put(4376,-3261){\makebox(0,0)[b]{\smash{\SetFigFont{10}{12.0}{\familydefault}{\mddefault}{\updefault}\special{ps: gsave 0 0 0 setrgbcolor}100\special{ps: grestore}
}}}
\put(4803,-3261){\makebox(0,0)[b]{\smash{\SetFigFont{10}{12.0}{\familydefault}{\mddefault}{\updefault}\special{ps: gsave 0 0 0 setrgbcolor}200\special{ps: grestore}
}}}
\put(5230,-3261){\makebox(0,0)[b]{\smash{\SetFigFont{10}{12.0}{\familydefault}{\mddefault}{\updefault}\special{ps: gsave 0 0 0 setrgbcolor}300\special{ps: grestore}
}}}
\put(5657,-3261){\makebox(0,0)[b]{\smash{\SetFigFont{10}{12.0}{\familydefault}{\mddefault}{\updefault}\special{ps: gsave 0 0 0 setrgbcolor}400\special{ps: grestore}
}}}
\put(6085,-3261){\makebox(0,0)[b]{\smash{\SetFigFont{10}{12.0}{\familydefault}{\mddefault}{\updefault}\special{ps: gsave 0 0 0 setrgbcolor}500\special{ps: grestore}
}}}
\put(6512,-3261){\makebox(0,0)[b]{\smash{\SetFigFont{10}{12.0}{\familydefault}{\mddefault}{\updefault}\special{ps: gsave 0 0 0 setrgbcolor}600\special{ps: grestore}
}}}
\put(5551,-3511){\makebox(0,0)[b]{\smash{\SetFigFont{10}{12.0}{\familydefault}{\mddefault}{\updefault}\special{ps: gsave 0 0 0 setrgbcolor}$\rm M_{1/2} \; ( GeV )$\special{ps: grestore}
}}}
\put(4151,-3115){\makebox(0,0)[b]{\smash{\SetFigFont{10}{12.0}{\familydefault}{\mddefault}{\updefault}\special{ps: gsave 0 0 0 setrgbcolor}100\special{ps: grestore}
}}}
\put(4151,-2830){\makebox(0,0)[b]{\smash{\SetFigFont{10}{12.0}{\familydefault}{\mddefault}{\updefault}\special{ps: gsave 0 0 0 setrgbcolor}150\special{ps: grestore}
}}}
\put(4151,-2546){\makebox(0,0)[b]{\smash{\SetFigFont{10}{12.0}{\familydefault}{\mddefault}{\updefault}\special{ps: gsave 0 0 0 setrgbcolor}200\special{ps: grestore}
}}}
\put(4151,-2261){\makebox(0,0)[b]{\smash{\SetFigFont{10}{12.0}{\familydefault}{\mddefault}{\updefault}\special{ps: gsave 0 0 0 setrgbcolor}250\special{ps: grestore}
}}}
\put(4151,-1976){\makebox(0,0)[b]{\smash{\SetFigFont{10}{12.0}{\familydefault}{\mddefault}{\updefault}\special{ps: gsave 0 0 0 setrgbcolor}300\special{ps: grestore}
}}}
\put(4151,-1691){\makebox(0,0)[b]{\smash{\SetFigFont{10}{12.0}{\familydefault}{\mddefault}{\updefault}\special{ps: gsave 0 0 0 setrgbcolor}350\special{ps: grestore}
}}}
\put(4151,-1407){\makebox(0,0)[b]{\smash{\SetFigFont{10}{12.0}{\familydefault}{\mddefault}{\updefault}\special{ps: gsave 0 0 0 setrgbcolor}400\special{ps: grestore}
}}}
\put(4151,-1122){\makebox(0,0)[b]{\smash{\SetFigFont{10}{12.0}{\familydefault}{\mddefault}{\updefault}\special{ps: gsave 0 0 0 setrgbcolor}450\special{ps: grestore}
}}}
\put(3901,-2118){\rotatebox{90.0}{\makebox(0,0)[b]{\smash{\SetFigFont{10}{12.0}{\familydefault}{\mddefault}{\updefault}\special{ps: gsave 0 0 0 setrgbcolor}$\mu \rm \; ( GeV )$\special{ps: grestore}
}}}}
\end{picture}